\documentclass[12pt,preprint]{aastex}
\usepackage{apjfonts}
\usepackage{epsfig}
\usepackage{amsmath}
\usepackage{graphicx}
\usepackage{color}
\usepackage{float}
\usepackage{morefloats}

\def\gsim{\lower 2pt \hbox{$\, \buildrel {\scriptstyle >}\over
{\scriptstyle \sim}\,$}}
\def\lsim{\lower 2pt \hbox{$\, \buildrel {\scriptstyle <}\over
{\scriptstyle \sim}\,$}}

\begin{document}

\title{BACKGROUND MODEL FOR THE LOW-ENERGY TELESCOPE OF \emph{INSIGHT-HXMT}}

\author{Jin-Yuan Liao\altaffilmark{1}, Shu Zhang\altaffilmark{1}, Yong Chen\altaffilmark{1}, Juan Zhang\altaffilmark{1}, Jing Jin\altaffilmark{1},
Zhi Chang\altaffilmark{1}, Yu-Peng Chen\altaffilmark{1}, Ming-Yu Ge\altaffilmark{1}, Cheng-Cheng Guo\altaffilmark{1,2}, Gang Li\altaffilmark{1}, 
Xiao-Bo Li\altaffilmark{1}, Fang-Jun Lu\altaffilmark{1}, Xue-Feng Lu\altaffilmark{1}, Jian-Yin Nie\altaffilmark{1}, Li-Ming Song\altaffilmark{1}, 
Yan-Ji Yang\altaffilmark{1}, Yuan You\altaffilmark{1,2}, Hai-Sheng Zhao\altaffilmark{1}, Shuang-Nan~Zhang\altaffilmark{1,2,3}}
\altaffiltext{1}{Key Laboratory of Particle Astrophysics, Institute of High Energy Physics,
Chinese Academy of Sciences, Beijing 100049, China; liaojinyuan@ihep.ac.cn}
\altaffiltext{2}{University of Chinese Academy of Sciences, Chinese Academy of Sciences, Beijing 100049, China}
\altaffiltext{3}{National Astronomical Observatories, Chinese Academy of Sciences, Beijing, 100012, China}

\begin{abstract}
With more than 150 blank sky observations at high Galactic latitude, we make a systematic study to the background of 
the Low Energy Telescope (LE) of the \emph{Hard X-ray Modulation Telescope} (dubbed as \emph{Insight-HXMT}). 
Both the on-ground simulation and the in-orbit observation indicate that the background spectrum mainly has two components. One is the 
particle background that dominates above 7~keV and its spectral shape is consistent in every geographical locations. Another is the diffuse X-ray 
background that dominates below 7~keV and has a stable spectrum less dependent of the sky region. The particle background spectral shape can be obtained from 
the blind detector data of all the blank sky observations, and the particle background intensity can be measured by the blind detector at $10-12.5$~keV.
The diffuse X-ray background in the high Galactic latitude can also be obtained from the blank sky spectra after 
subtracting the particle background. Based on these characteristics, we develop the background model for both the spectrum 
and the light curve. The systematic error for the background spectrum is investigated with different exposures ($T_{\rm exp}$). 
For the spectrum with $T_{\rm exp}=1$~ks, the average systematic errors in $1-7$~keV and $1-10$~keV are $4.2\%$ and $3.7\%$, respectively.
We also perform the systematic error analyses of the background light curves with different energy bands and time bins. 
The results show that the systematic errors for the light curves with different time bins are $<8\%$ in $1-10$~keV. 
\end{abstract}

\keywords{instrumentation: detectors --- space vehicles: instruments --- methods: data analysis --- X-rays: general}

\section{INTRODUCTION}
\begin{figure*}
\center{
\includegraphics[angle=0,scale=0.36]{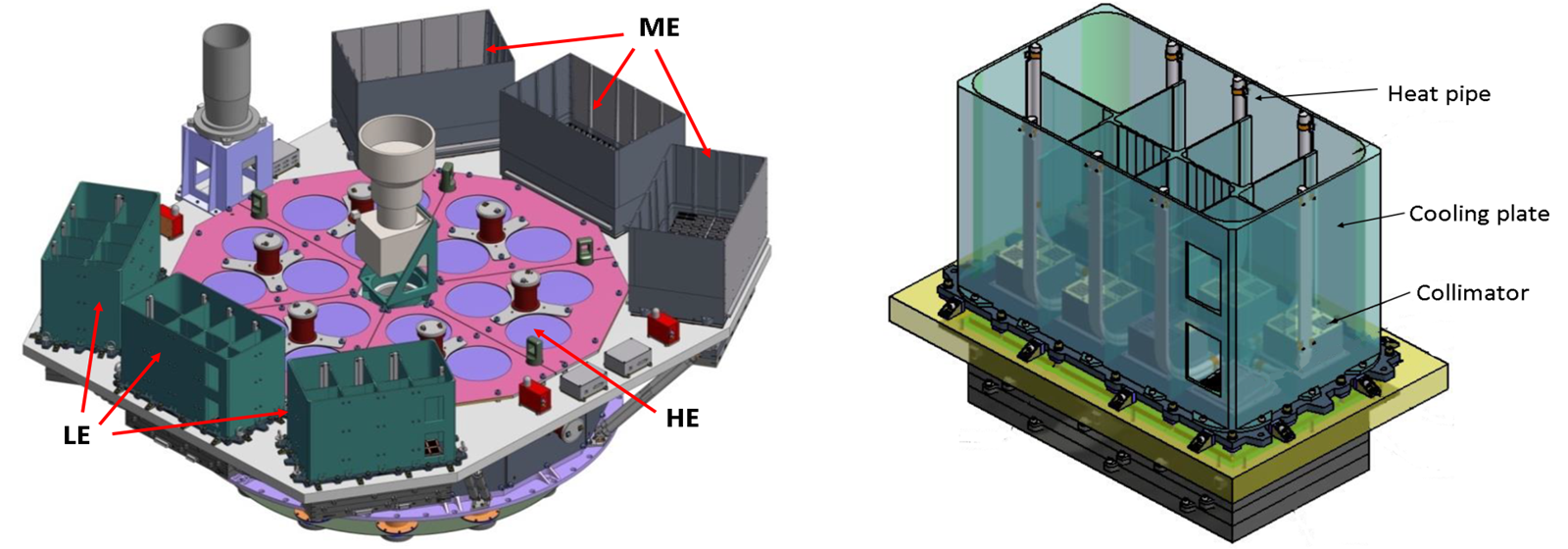}}
\caption{The main structrue of \emph{Insight-HXMT} and the LE detector modules.}
\label{Fig:HXMT_LE_structrue}
\end{figure*}

\begin{figure*}
\center{
\includegraphics[angle=0,scale=1.00]{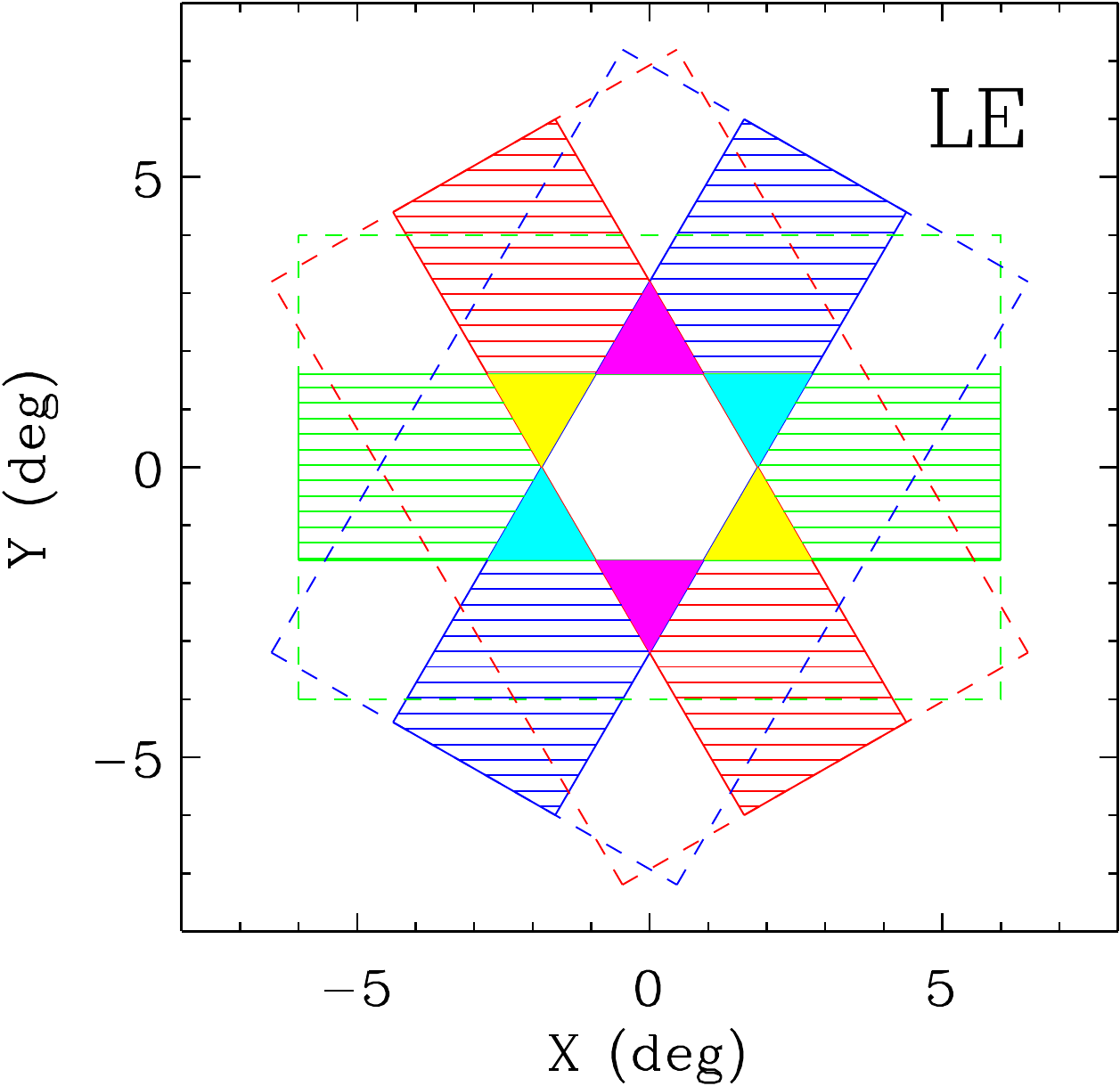}}
\caption{FOVs of the \emph{Insight-HXMT}/LE detectors. Three rectangular shadows shown in red, blue and green are three small FOVs
(FWHM: $1^\circ.6\times6^\circ$); and the three rectangles surrounded by the dashed lines are three large FOVs (FWHM: $4^\circ\times6^\circ$).}
\label{Fig:FOV_LE}
\end{figure*}

The Low Energy X-ray Telescope (LE) is one of the three main payloads of the \emph{Hard X-Ray Modulation Telescope} (dubbed as \emph{Insight-HXMT}) 
that has been operating in orbit since June 15th, 2017 (\citealt{2020_Zhang_SCPMA}). LE is composed of Swept Charge Devices (SCDs) that are sensitive 
in $0.7-13$~keV with a total geometrical area of 384~cm$^2$ (\citealt{2020_Chen_SCPMA}). By using LE and the other two main payloads together, i.e., 
the Medium Energy X-ray Telescope ($5-30$~keV, \citealt{2020_Cao_SCPMA}) and the High Energy X-ray Telescope ($20-250$~keV, \citealt{2020_Liu_SCPMA}),
\emph{Insight-HXMT} can be used for wide-band spectral and temporal analyses. 
The main structrue of \emph{Insight-HXMT} and the LE detector modules are shown in Figure~\ref{Fig:HXMT_LE_structrue}.
As described in \citet{2020_Chen_SCPMA}, LE is a collimated type of telescope that consists of three detector boxes with orientation of the field of views (FOVs) 
offset by $60^\circ$ (Figure~\ref{Fig:FOV_LE}). Each detector box contain 20 SCD type of detectors with the small FOVs ($1^\circ.6\times6^\circ$), 
six detectors with the large FOVs ($4^\circ\times6^\circ$), one blind detector with a small FOV collimator, and one blind detector with a large FOV 
collimator that is accompanied by a Fe$^{55}$ radioisotope to monitor the possible change of the energy response.

\emph{Insight-HXMT} operates on a quasi-circular low-earth orbit with an attitude of $\sim550$~km and an inclination of $\sim43^\circ$. 
There are various particles that can interact with the satellite platform and the payloads, e.g., the cosmic-ray protons and electrons 
(Alcaraz et al. 2000a, 2000b), the gamma-rays and neutrons reflected from the earth (\citealt{1976JGR....81.2835I}; \citealt{1973JGR....78.2715A}), 
and the cosmic X-ray background (CXB, \citealt{1992NIMPA.313..513G}, \citealt{2010PhRvL.104j1101A}). The complexity of the space environment is also 
revealed by long-term observations of the particle monitor onboard \emph{Insight-HXMT} (\citealt{2019_Lu_JHEAp}; \citealt{2020_Liao_JHEAp_HE}).
According to the on-ground simulation (\citealt{2015_Li_PIA}), the LE background can be roughly divided into two parts, i.e., the particle background 
that is dominant in high energy band ($>7$~keV) and the diffuse X-ray background that is dominant in high energy band ($<7$~keV).

In this paper, we focus on the particle background and the diffuse X-ray background in high Galactic latitude ($\left|b\right|<10^\circ$) 
and the region far from the Galactic center ($90^\circ<l<270^\circ$). 
The diffuse X-ray background in the Galactic center is strong and complex (\citealt{1997ApJ...485..125S}).
LE has relatively large FOVs, thus the diffuse X-ray background of LE in the Galactic center needs to be considered seriously. 
Currently, the Galactic diffuse background is estimated from \emph{Insight-HXMT} Galactic plane scanning survey,
for which the details will be reported in a forthcoming paper. 


This paper is organized as follows. The data reduction of \emph{Insight-HXMT}/LE is shown in Section 2. In Section 3, we show the main 
observational characteristics of the background of \emph{Insight-HXMT}/LE. In Section 4, we describe the method to estimate the LE background. 
The model test and discussion of the background estimation are shown in Section 5, respectively. The summary is given in Sections 6.

\section{DATA REDUCTION}
\subsection{Preliminary data reduction}
The preliminary data reduction is performed with the \emph{Insight-HXMT} data analysis software HXHTDAS v2.1. 
For every blank sky observation, three tools are run before the spectra and the light curves are generated, as shown in the follows.
\begin{itemize}
    \item lepical: PI transformation to obtain the energy-calibrated events..
    \item lerecon: Event reconstruction to select the non-split events.
    \item legtigen: Good Time Interval (GTI) selection to obtain the data within the specified parameter range. 
\end{itemize} 
In LE GTI selection, the earth elevations (ELV) are selected as $>10^\circ$ to avoid the earth occultation of the blank sky. 
As the LE backgrounds are usually unstable due to the complex space environment near the South Atlantic Anomaly (SAA), 
the time since the last passage from SAA ($\rm T\_SAA$) and the time to the next SAA passage ($\rm TN\_SAA$) are both set to $>300$~s. 
The geomagnetic cut-off rigidity (COR) is anti-correlated to particle flux and the LE background level. Since the LE background 
is high but stable in the low COR region, COR is ignored in the LE GTI selection. When a large number of the visible light or 
low-energy charged particles enter the collimator, the detectors will be saturated, thus $\rm DYE\_ELV$ is generally chosen as $>20^\circ$ 
in the regular pointing observations. However, there are still some anomalous peaks in the light curve when the satellite
passes through the precipitating particle areas near the equator and the high latitude regions, even if $\rm DYE\_ELV$ is $>60^\circ$. 
Therefore, $\rm DYE\_ELV$ is also ignored in the LE GTI selection and a more reliable and 
efficient method is added to the LE GTI selection instead.

\subsection{GTI prior selection}
\begin{figure*}
\center{
\includegraphics[angle=0,scale=1.00]{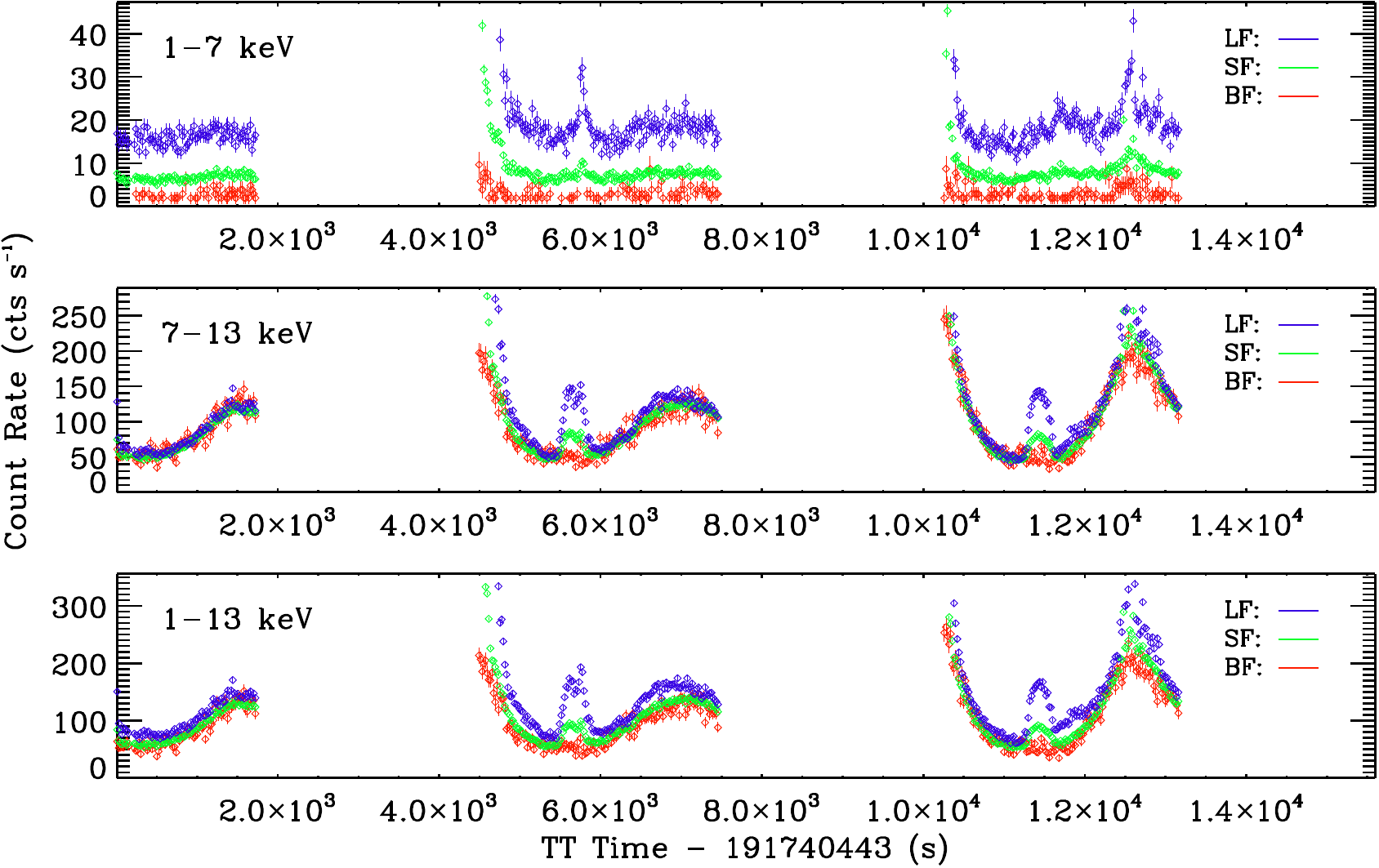}}
\caption{Light curves of a blank sky observation of \emph{Insight-HXMT}/LE in $1-7$~keV (top), $7-13$~keV (middle), and $1-13$~keV (bottom).
In all three panels, the light curve of blind detector ($L_{\rm Blind}$), small FOV detector ($L_{\rm Small}$) and large FOV detector ($L_{\rm Large}$) 
are shown in red, green and blue, respectively.
Both the light curves of the blind and large FOV detectors are normalized by the ratio of the detector number with the small FOV  detectors.}
\label{Fig:LC_3_FOV}
\end{figure*}

\begin{figure*}
\center{
\includegraphics[angle=0,scale=1.00]{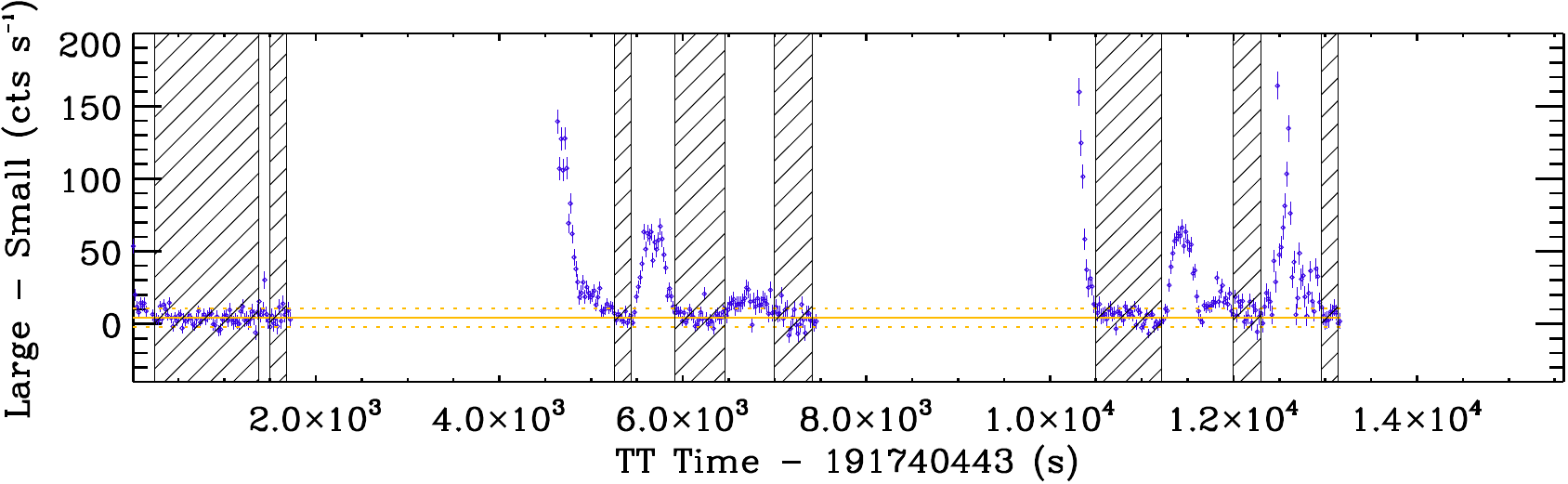}}
\caption{Difference between the light curves of large and small FOV detectors in $7-13$~keV shown in Figure~\ref{Fig:LC_3_FOV}.
The time after the GTI prior selection is marked with shadows.}
\label{Fig:Large-Small}
\end{figure*}

For a pointing observation with the normal detector state, the variabilities of the source and the particle background are similar 
among the detectors with the small and large FOVs. Moreover, the diffuse X-ray backgrounds are proportional to the FOVs and can be 
considered as a constant for each detector. Therefore, the difference between the light curve of the large and small FOV detectors is 
almost constant since they are caused by the different contribution of the diffuse X-ray backgrounds. 
Figure~\ref{Fig:LC_3_FOV} shows an example of a LE light curve of a blank sky observation, where several flares are obvious in the 
light curves of both the large and small FOVs. Compared to the low energy band ($1-7$~keV), the flares are much more significant 
in the high energy band ($7-13$~keV). As the flares are proportional to the FOVs and have a typical duration of several hundreds seconds, which is very 
different from the diffuse X-ray backgrounds, we speculate that the flares may be the result of a large number of low-energy charged particles entering 
the collimator. Such a speculation is further supported by the on-ground Geant4 simulation. As shown in Figure~\ref{Fig:Large-Small}, 
the difference of the two light curves in $7-13$~keV remains constant for most of the time, hence it serves as a prior GTI selection criterion for
removing the flares events, which is described in what follows in details.
\begin{itemize}
    \item Obtain the difference of the light curves of the large and small FOV detectors.
	\item The time period with the difference remains constant for more than 100~s (to avoid accidents) is adopted as the GTI of \emph{Insight-HXMT}/LE. 
 \end{itemize}
For the pointing observation, the GTI prior selection is made first, and then the regular GTI selection is performed to obtain the final GTI of \emph{Insight-HXMT}/LE.

\section{OBSERVATIONAL CHARACTERISTICS OF LE BACKGROUND}
\begin{figure*}
\center{
\includegraphics[angle=0,scale=1.00]{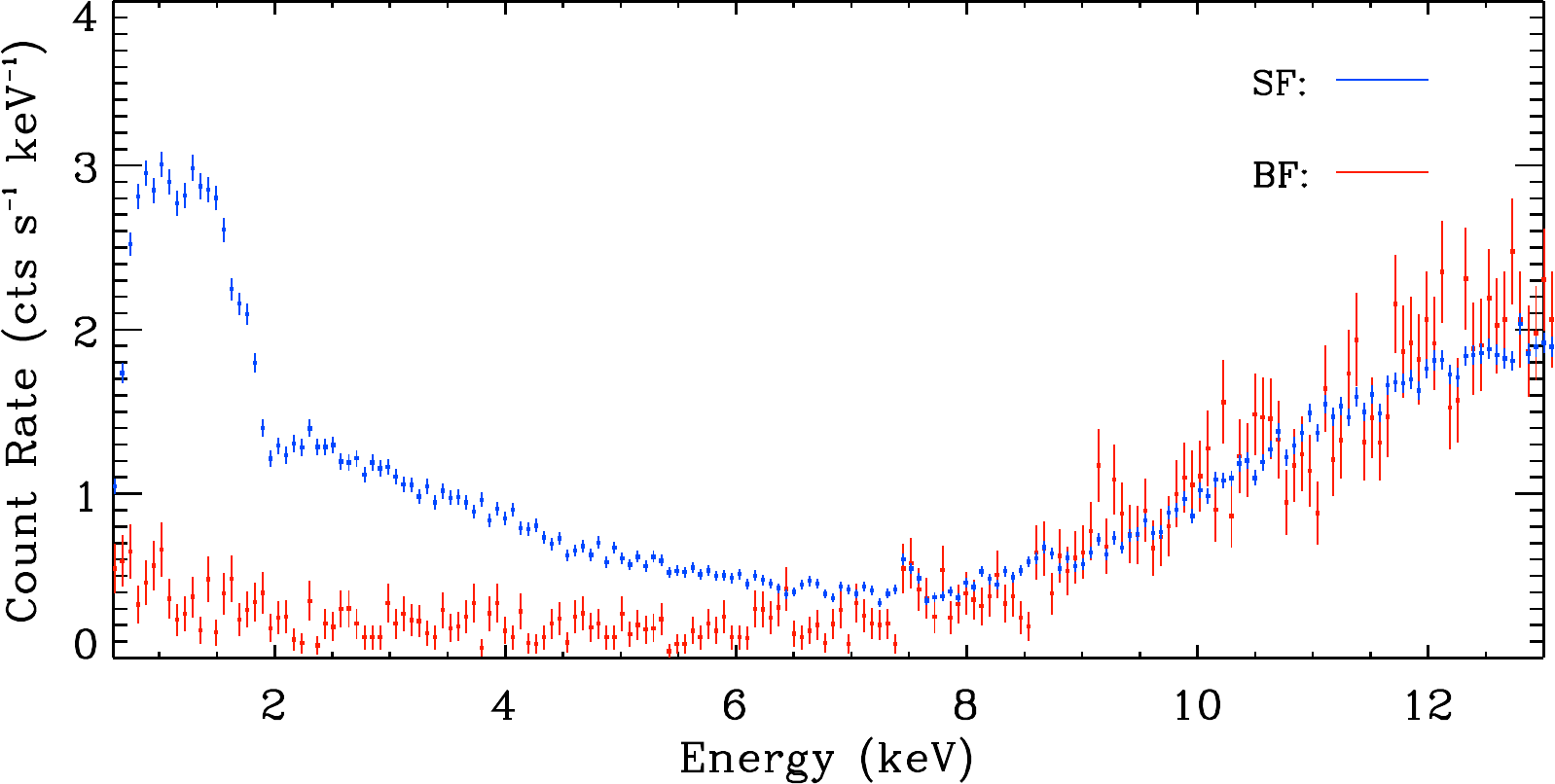}}
\caption{Spectra of a blank sky observed with the small FOV detector (blue) and the blind FOV detector (red).}
\label{Fig:le_spec}
\end{figure*}

\begin{figure*}
\center{
\includegraphics[angle=0,scale=1.00]{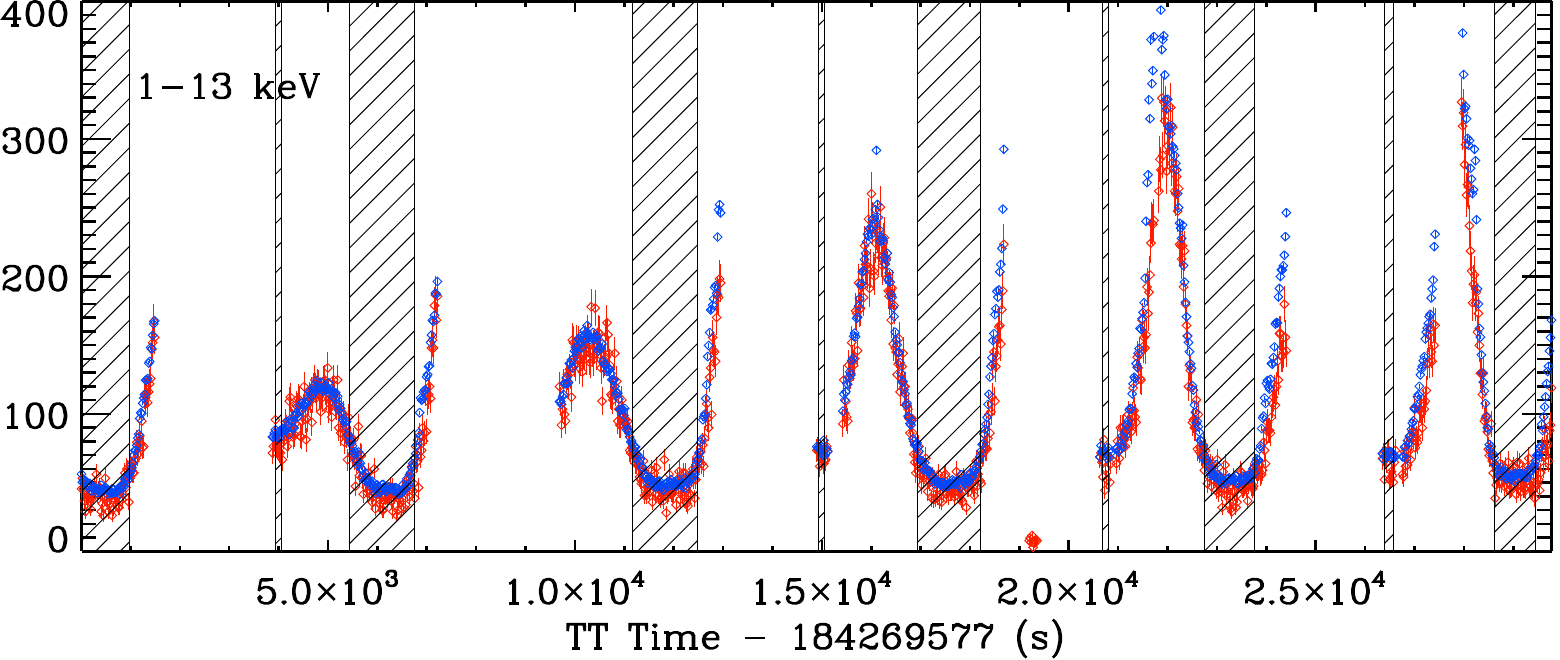}}
\caption{Light curves of a blank sky observed with the small FOV detector (blue) and the blind FOV detector (red).
The final GTI is marked with shadows.}
\label{Fig:le_lc}
\end{figure*}

\begin{figure*}
\center{
\includegraphics[angle=0,scale=1.00]{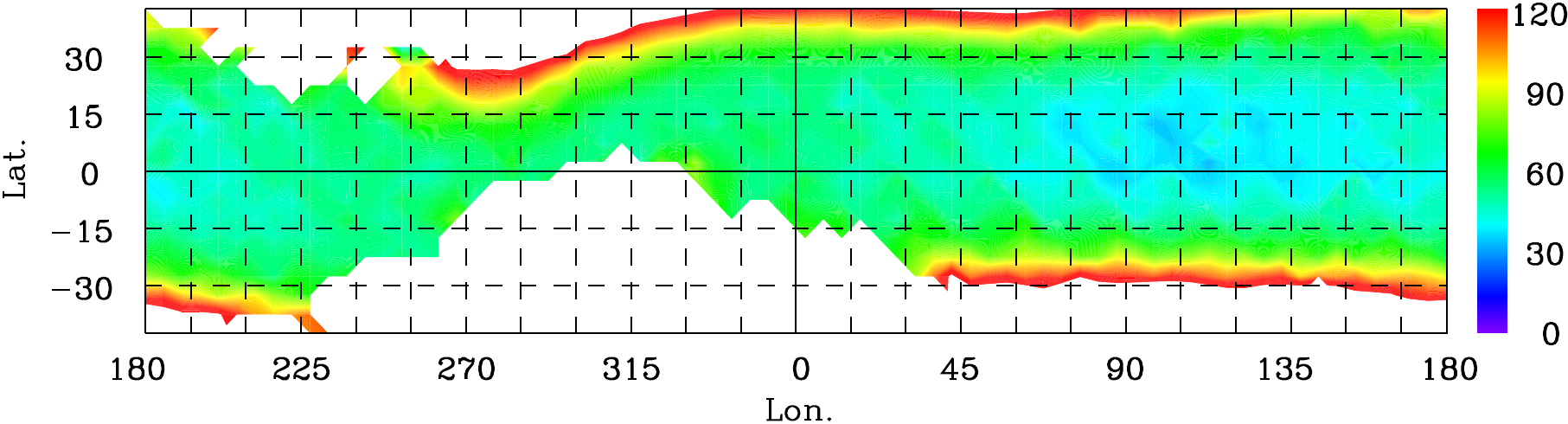}}
\caption{Geographical distribution of the background intensity (small FOV detector \& $1-13$~keV).}
\label{Fig:bkg_map}
\end{figure*}

Figure~\ref{Fig:le_spec} shows the spectra of a blank sky observation with the blind and small FOV detectors of \emph{Insight-HXMT}/LE.
Because an aluminum plate blocks 
the FOV of the collimator, the blind detector cannot receive the X-ray photons in the LE working band, thus the blind detector can be 
considered as the pure particle background spectrum estimator (shown with the red color in Figure~\ref{Fig:le_spec}). The background spectra 
of the small FOV detectors above 7~keV are consistent with these of the blind detectors, and the diffuse X-ray backgrounds are dominant 
below 7~keV (shown with the blue color in Figure~\ref{Fig:le_spec}). Therefore, the particle background spectra of the small FOV detectors 
can be indicated with the spectra of the blind detectors, and the difference between the spectra of the blind and small FOV detectors is the 
diffuse X-ray background detected by \emph{Insight-HXMT}/LE. An obvious geographical modulation shows up in the light curve (Figure~\ref{Fig:le_lc}),
due to the anti-correlation between the particle background and COR (Figure~\ref{Fig:bkg_map}). Such a phenomenon is indicated in the on-ground 
simulation (\citealt{2015_Li_PIA}) and observed by other telescopes (e.g., \emph{Suzaku}, \citealt{2009PASJ...61S..17F}).

\section{MODELING OF LE PARTICLE BACKGROUND}
\subsection{Spectrum of LE Particle Background}
The background of LE is mainly composed of several prompt background components. In high-energy band ($>7$~keV), 
the particle background is dominant with intensity varying with the geographical locations; and in low-energy band ($<7$~keV), 
the diffuse X-ray background is dominant with intensity remaining constant during a pointing observation.

\begin{figure*}
\center{
\includegraphics[angle=0,scale=1.00]{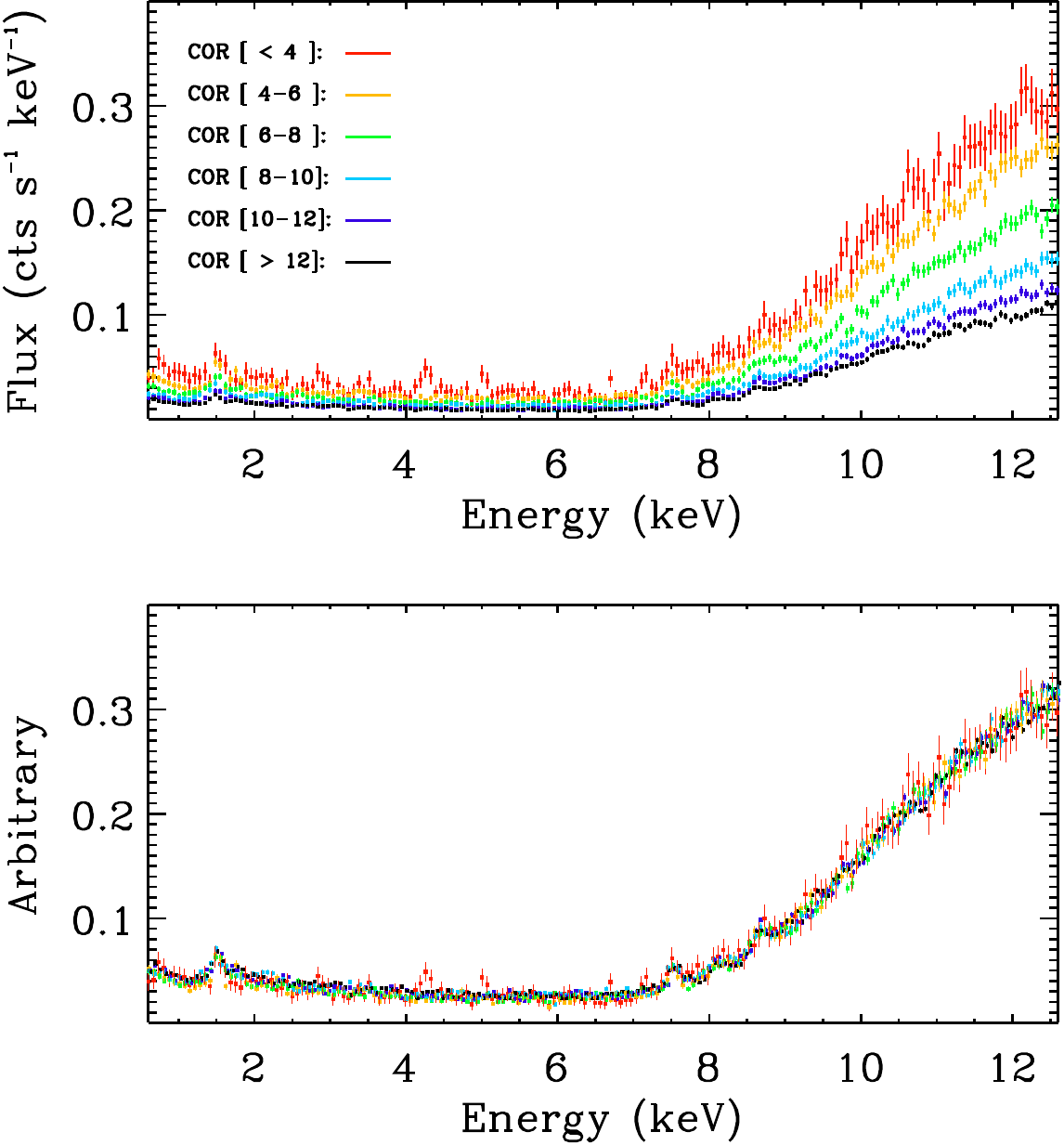}}
\caption{Top: Spectra of the blind FOV detector with different COR ranges. 
Bottom: normalized spectra of the top panel.}
\label{Fig:spec_cor}
\end{figure*}

\begin{figure*}
\center{
\includegraphics[angle=0,scale=1.00]{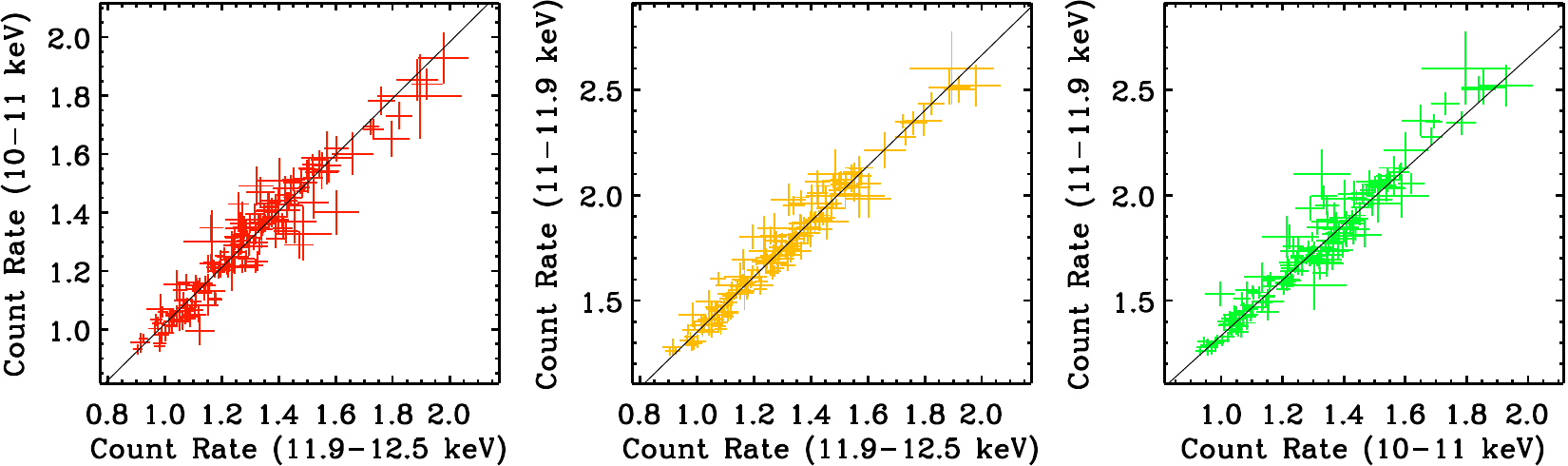}}
\caption{Correlations of the count rates in 10--11~keV, 11--11.9~keV and 11.9--12.5~keV. 
All the data are obtained from the blank sky observations with the small FOV detector.}
\label{Fig:spec_index}
\end{figure*}

\begin{figure*}
\center{
\includegraphics[angle=0,scale=1.00]{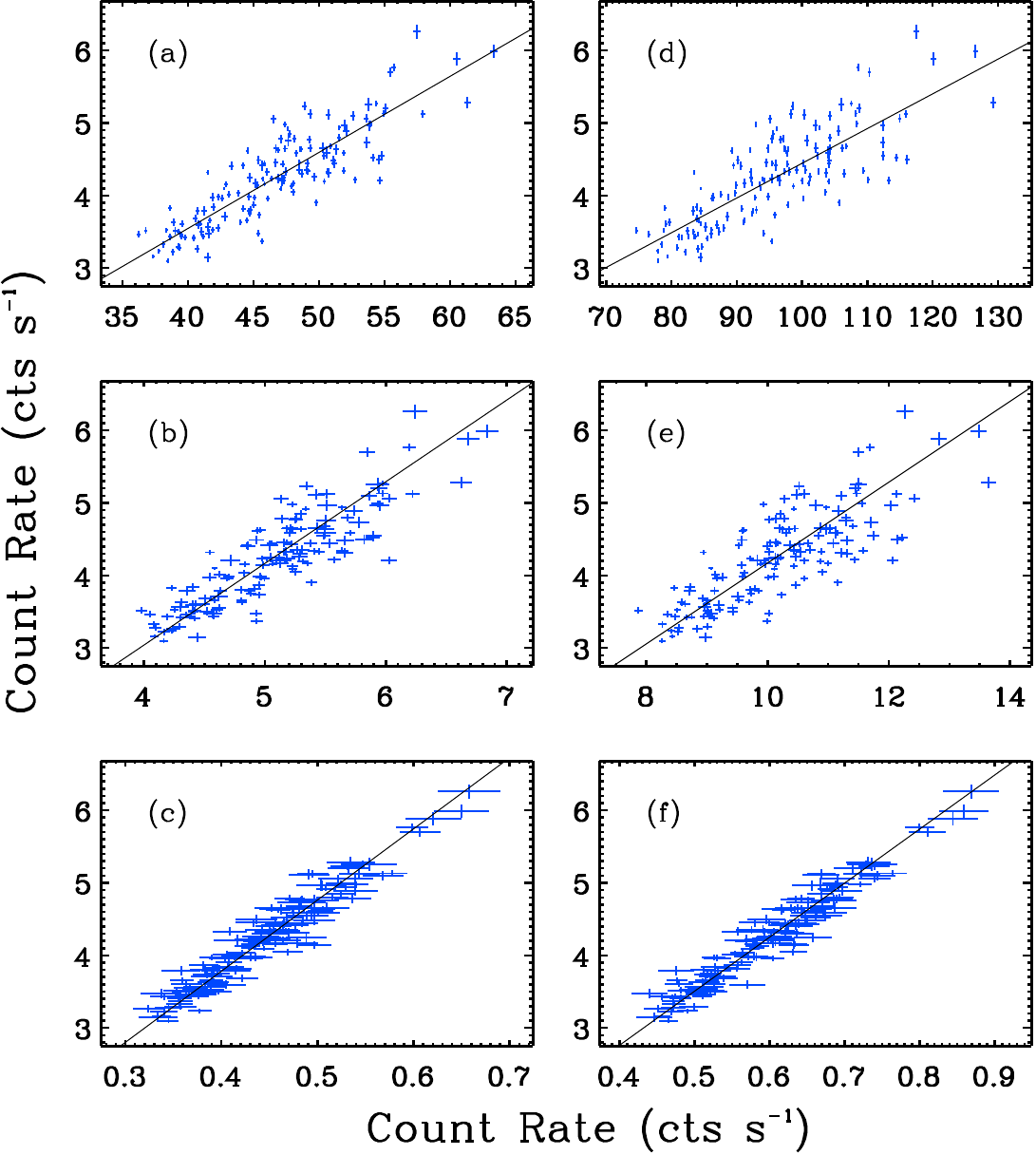}}
\caption{Correlation between the particle background of the small FOV detector and different variables.
Panel (a): the count rate of the single event exceeding the upper threshold of the small FOV detector;
Panel (b): the count rate of the single event exceeding the upper threshold of the blind FOV detector;
Panel (c): the count rate of the single event in $10-12.5$~keV of the blind FOV detector;
Panel (d)--(f): the same as Panel (a)--(c) but also with the splitting events.
}
\label{Fig:spec_correlation}
\end{figure*}

Based on more than 150 observations of the blank sky in high Galactic latitude, we find that the spectral shapes of the 
particle backgrounds are stable in different COR ranges for both the blind and small FOV detectors (Figure~\ref{Fig:spec_cor}--\ref{Fig:spec_index}). 
Accordingly, the estimation of the background spectra is highly be simplified: the spectrum of the particle background is measured first 
and then the background model is completed by adjusting the intensity. All we need to do is to find the best indicator of the 
background intensity. A variety of parameter combinations are investigated (Figure~\ref{Fig:spec_correlation}), 
which includes but are not limited to the signals over the upper threshold (with and without the splitting events), 
the integrated count rate of the blind detector in high-energy band (with and without the splitting events). 
By investigating the the correlation between different parameters, we find that the integrated count rate ($10-12.5$~keV) 
is mostly correlated to the background intensity (Figure~\ref{Fig:spec_correlation}c). In order to reduce the statistical errors, 
both the blind detectors with small and large FOVs are used, i.e., the integrated count rate ($10-12.5$~keV) of the small and large FOVs detectors are used to indicate the background intensity. The details are shown as follows.

a. Analyze every blank sky observation and merge all the spectra of the blind detectors, used as the spectral shape 
of the background of the small FOV detectors: $S_{\rm Mod}(c|{\rm SD})$

b. Analyze every blank sky observation and obtain the count rate in $10-12.5$~keV ($\Delta E_{\rm h}$) of both the detectors with small and large FOVs: 
$R_{{\rm BS},i}(\Delta E_{\rm h}|{\rm SD})$ and $R_{{\rm BS},i}(\Delta E_{\rm h}|{\rm BD})$, where BS denote the ``Blank sky'' and $i=0,1,...,n$. 

c. Obtain the ratio of the count rate in $10-12.5$~keV between the small and blind FOV detectors 
\begin{equation}
f = \frac{\sum_{\rm All}R_{{\rm BS},i}(\Delta E_{\rm h}|{\rm SD})}{\sum_{\rm All}R_{{\rm BS},i}(\Delta E_{\rm h}|{\rm BD})},
\end{equation}

d. For a pointing observation, with the count rate of the blind FOV detector in $10-12.5$~keV $R_{\rm Obs}(10-12.5~{\rm keV|BD})$ and the ratio in step c, obtain the expected count rate of the particle background of the small FOV detector in $10-12.5$~keV
\begin{equation}
R_{\rm Exp}(\Delta E_{\rm h}|{\rm SD}) = R_{\rm Obs}(\Delta E_{\rm h}|{\rm BD}) \cdot f,
\end{equation}

e. Obtain the correction factor of the background intensity
\begin{equation}
F = \frac{R_{\rm Exp}(\Delta E_{\rm h}|{\rm SD})}{R_{\rm Mod}(\Delta E_{\rm h}|{\rm SD})},
\end{equation}

f. Make the correction
\begin{equation}
S_{\rm Est}(c|{\rm SD}) = F \cdot S_{\rm Mod}(c|{\rm SD}).
\end{equation}
The whole calculation process of the background spectra can be described as
\begin{equation}
S_{\rm Est}(c|{\rm SD}) = \frac{R_{\rm OBS}(\Delta E_{\rm h}|{\rm BD})}{R_{\rm Mod}(\Delta E_{\rm h}|{\rm SD})} 
\frac{\sum_{\rm All}R_{{\rm BS},i}(\Delta E_{\rm h}|{\rm SD})}{\sum_{\rm All}R_{{\rm BS},i}(\Delta E_{\rm h}|{\rm BD})} S_{\rm Mod}(c|{\rm SD}).
\end{equation}

\subsection{Light Curve of LE Particle Background}
The background model for light curve follows a procedure similar to that for energy spectrum and is shown in details in what follows.

a. Obtain the observational light curve of the blind FOV detectors in $10-12.5$~keV: $L_{\rm Obs}(t;\Delta E_{\rm h}|{\rm BD})$.

b. Obtain the expected background light curve of the small FOV detector in $10-12.5$~keV
\begin{equation}
L_{\rm Exp}(t;\Delta E_{\rm h}|{\rm SD}) = L_{\rm Obs}(t;\Delta E_{\rm h}|{\rm BD}) f,
\end{equation}

c. Based on the background characteristics that the spectral shape is consistent in any geographical locations, obtain the ratio of the 
expected background in the special energy band and $10-12.5$~keV
\begin{equation}
\mathcal{R}(t;\Delta E) = \frac{S_{\rm Mod}(\Delta E|{\rm SD})}{S_{\rm Mod}(\Delta E_{\rm h}|{\rm SD})},
\end{equation}

d. Finally, the background light curve of the special energy band can be obtained by
\begin{equation}
\begin{aligned}
L_{\rm Est}(t;\Delta E|{\rm SD}) & = L_{\rm Exp}(t;\Delta E_{\rm h}|{\rm SD}) \mathcal{R}(t;\Delta E) \\
          & = L_{\rm Obs}(t;\Delta E_{\rm h}|{\rm BD}) \frac{S_{\rm Mod}(\Delta E|{\rm SD})}{S_{\rm Mod}(\Delta E_{\rm h}|{\rm SD})} f.
\end{aligned}
\end{equation}

In order to reduce the statistical error, the background light curve with $T_{\rm bin} = 16~{\rm s}$ is produced first.
In addition, both the on-ground simulation and the in-orbit observation indicate the time scale of the variability of the 
LE background is always very long ($>100$~s). Thus the background light curve with shorter time bin can be obtained by 
interpolating the background light curve with $T_{\rm bin} = 16~{\rm s}$.

\section{MODEL TEST AND DISCUSSION}
\subsection{Estimation of the Diffuse Background in High Galactic Latitude}
The observations of 21 blank sky regions in high Galactic latitudes are used to test the background model.
For the blank sky observations in high Galactic latitude, the diffuse X-ray background caused by the CXB in the low energy band 
dominates the LE background spectrum. 

\begin{figure*}
\center{
\includegraphics[angle=0,scale=1.00]{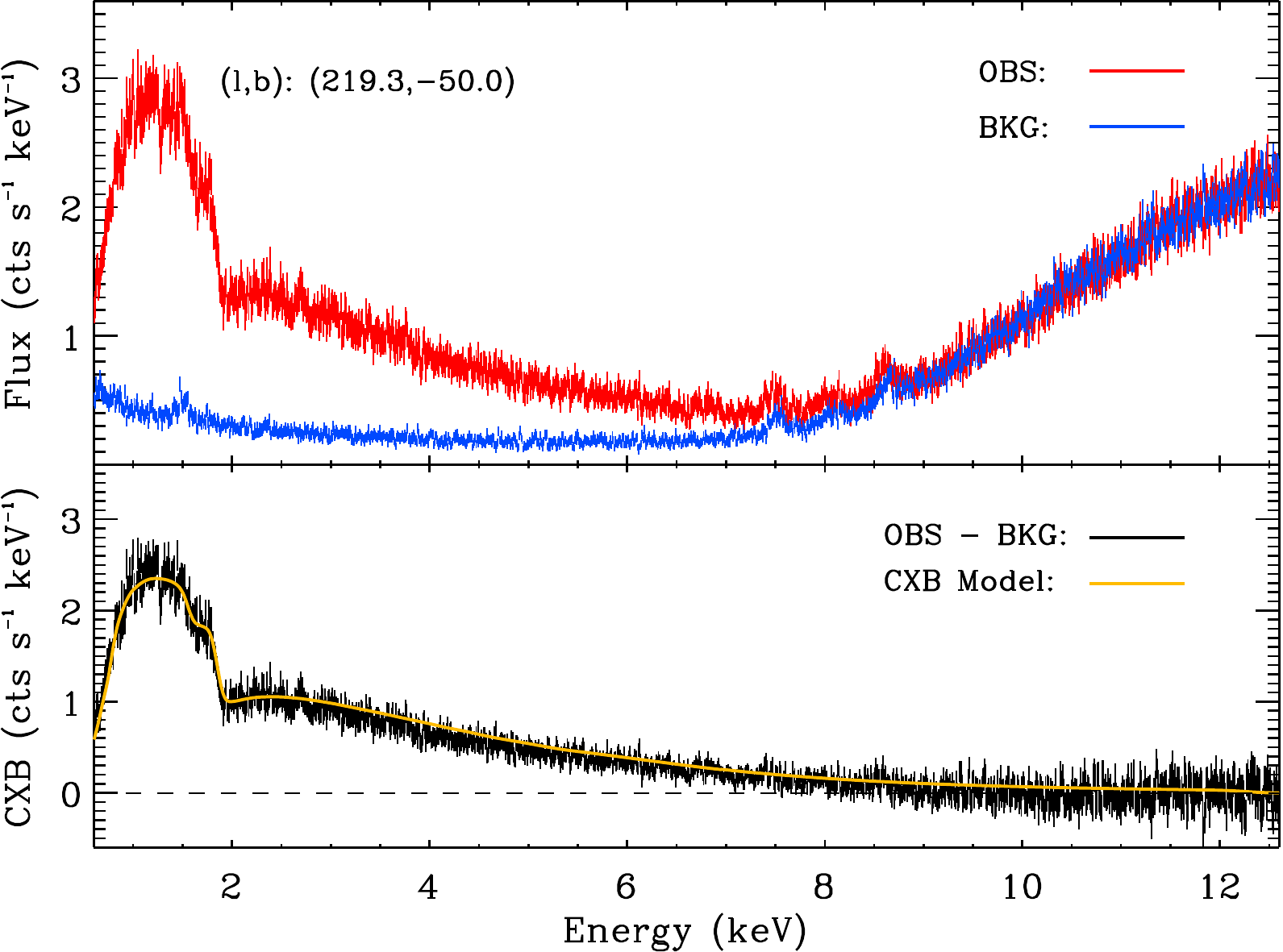}}
\caption{Top: spectrum of a blank sky observation (red) and the particle background spectrum (blue). 
Bottom: the CXB spectrum obtained from the blank sky observation by subtracting the particle background;
The yellow curve is the model with the CXB spectral parameters (\citealt{2003A&A...411..329R}) and the response of \emph{Insight-HXMT}/LE.}
\label{Fig:cxb_hxmt_rxte}
\end{figure*}

\begin{figure*}
\center{
\includegraphics[angle=0,scale=1.00]{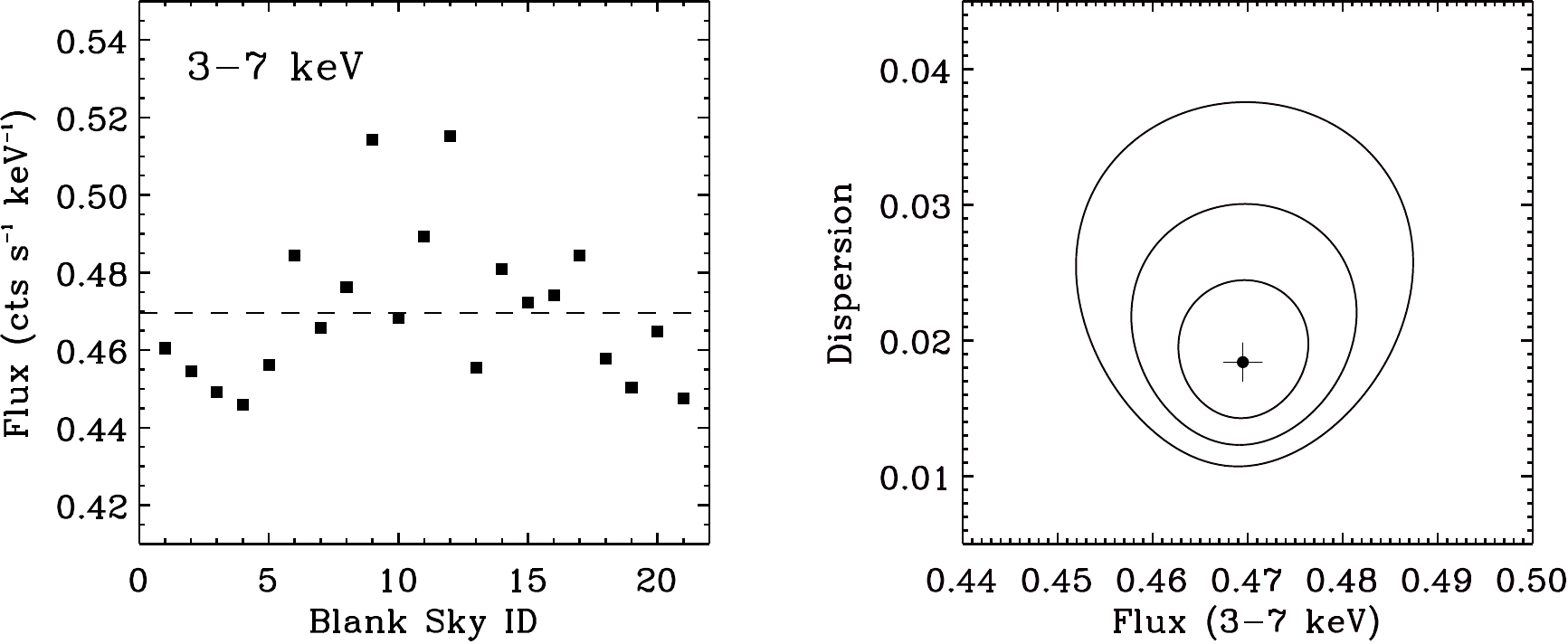}}
\caption{Left: the CXB count rates of 21 blank sky regions observed with \emph{Insight-HXMT} in 3-7 keV. 
Right: the 2D posterior distribution of the mean value and the intrinsic dispersion of the CXB count rates obtained from the Bayesian analysis of the data in the left panel; 
the cross marks the maximum a posteriori estimates and the three contours from the inside out are the 1$\sigma$, 2$\sigma$ and 3$\sigma$ credible intervals, respectively.}
\label{Fig:cxb_analysis}
\end{figure*}

Therefore, in order to test the particle background model, the CXB spectra must be obtained first. 
For every blank sky, the particle background spectrum is obtained with the background model first 
and then subtracted from the observational spectrum, thus the observed CXB spectrum with 
\emph{Insight-HXMT} is obtained (Figure~\ref{Fig:cxb_hxmt_rxte}). 
With the maximum likelihood estimation described in \citet{2013ApJ...774..116L}, 
we obtain the intrinsic dispersion of the CXB count rates in $3-7$~keV among the adopted blank sky regions is $4.0\pm0.7\%$ (Figure~\ref{Fig:cxb_analysis}). 
As shown in \citet{2003A&A...411..329R}, CXB has a fluctuation of ~7\% 
per square degree, and with the LE small FOV $1^\circ.6\times6^\circ$, the expected fluctuation of the LE diffuse X-ray background is $\sim2.3\%$. 
Therefore, the measured CXB fluctuation of LE is slightly larger than the theoretical expectation, 
which may be due to the systematic error of calibration and the unresolved sources in each blank sky. 
However, such a deviation is relatively small and insignificant. 
Moreover, the CXB spectra obtained by \emph{Insight-HXMT} can be well matched by the CXB spectral model with
parameters given by \emph{RTXE}/PCA (\citealt{2003A&A...411..329R}), which indicates that the particle background is reliable.
In the following, the average of all the diffuse X-ray background spectra 
is considered as the diffuse X-ray background of a pointing observation to the high Galactic latitude region. 
We take the diffuse X-ray background in the high Galactic latitude as input and combine it with the particle background model to estimate 
the systematic error of the LE background modeling.

\subsection{Test for the Background Spectra}
\begin{figure*}
\center{
\includegraphics[angle=0,scale=1.00]{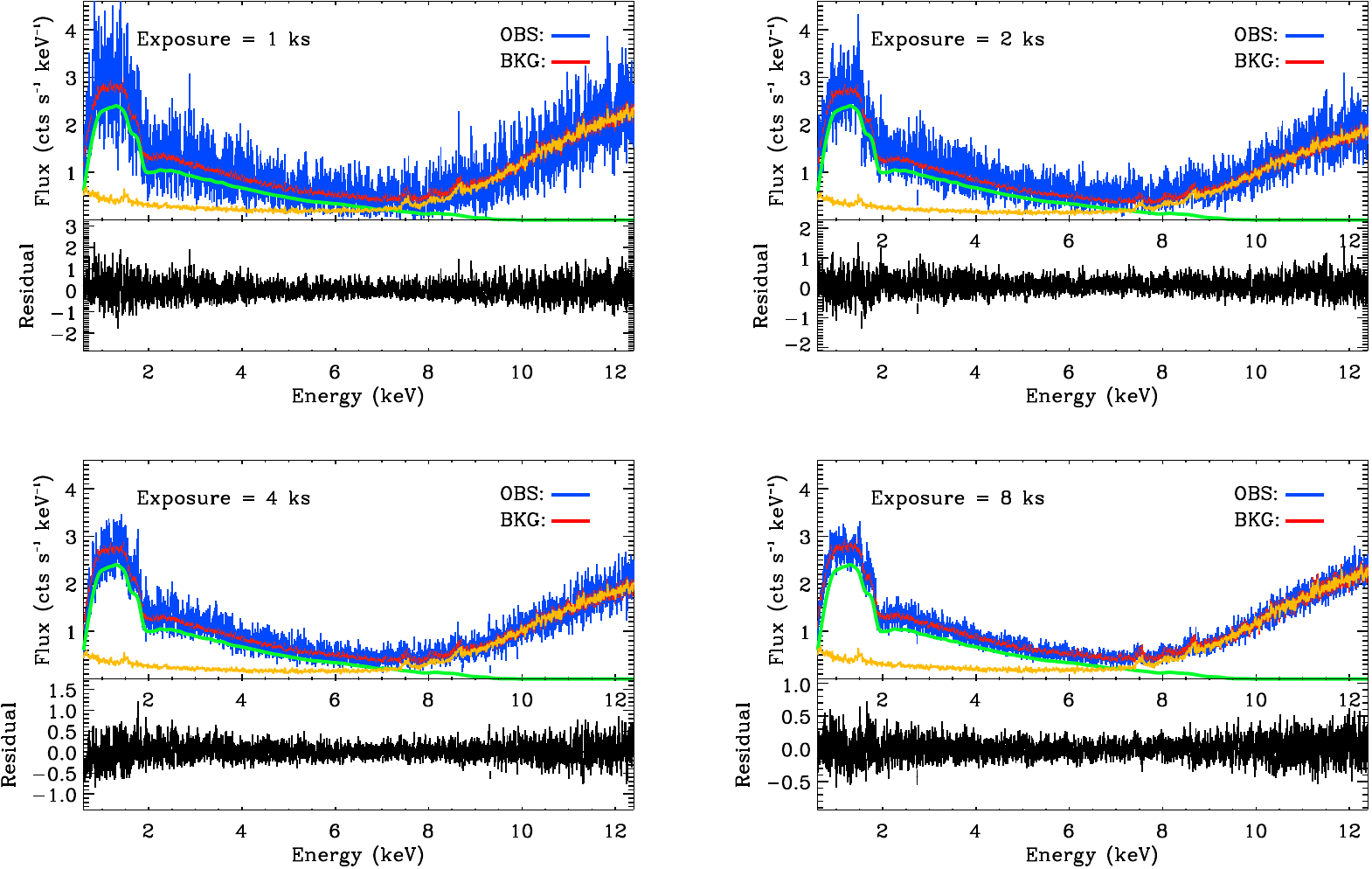}}
\caption{Background spectra test with the exposures 1~ks, 2~ks, 4~ks and 8~ks. 
Top of each panel: blue is the observed spectrum of a blank sky, red is the estimated background spectrum 
that is composed of the CXB spectrum (green) and the particle background spectrum (yellow).
Bottom of each panel: residual of the background spectra estimation.}
\label{Fig:bkg_spec_check}
\end{figure*}

In order to estimate the exposure dependence of the background spectrum model,
all the blank sky observations are re-grouped into sub-observations with the same exposure ($T_{\rm exp}$).
Figure~\ref{Fig:bkg_spec_check} shows the background spectral estimation of a blank sky observation with $T_{\rm exp} = 1$~ks, 
2~ks, 4~ks and 8~ks, respectively. An example of testing the background spectrum with $T_{\rm exp} = 1$~ks is shown as follows.

(1) Re-group all the blank sky observations into $N$ sub-observations with $T_{\rm exp} = 1$~ks
and obtain their observational spectra $S_{i,\rm obs}(c)$, where $i=1,2,...,N$.

(2) Obtain the particle background spectra of these sub-observations with the particle background model.

(3) Combine the particle background spectra and the diffuse X-ray background spectra
to obtain the total background spectra $S_{i,\rm bkg}(c)$, where $i=1,2,...,N$.

(4) Compare $S_{\rm obs}(c)$ and $S_{\rm bkg}(c)$ to obtain the residuals $R(c) = S_{\rm obs}(c) - S_{\rm bkg}(c)$.


(5) Calculate the mean value and intrinsic dispersion of $R(c)$ of each channel, i.e., $R_{\rm m}$ and $D_{\rm in}$, by solving the following equation
\begin{equation}
\sum_{i=1}^N \frac{(R_{i}-R_{\rm m})^2}{D_{i}^2} = N - 1,
\end{equation}
where
\begin{equation}
    D_{i}^{2} = D_{\rm in}^{2} + \sigma_{i}^{2},
\end{equation}
\begin{equation}
    R_{\rm m} = \sum_{i=1}^N R_{i}\times w_{i}, \quad w_{i} = \frac{\frac{1}{D_{i}^{2}}}{\sum_{i=1}^N \frac{1}{D_{i}^{2}}},
\end{equation}
where $R_i$ refers to the $R(c)$ of every observation, $\sigma_{i}$ the statistic errors of $R_i$. 
$R_{\rm m}$ and $D_{\rm in}$ can be considered as the bias ($B$) and the systematic error ($\sigma_{\rm sys}$) of the background estimation.

\begin{figure*}
\center{
\includegraphics[angle=0,scale=1.00]{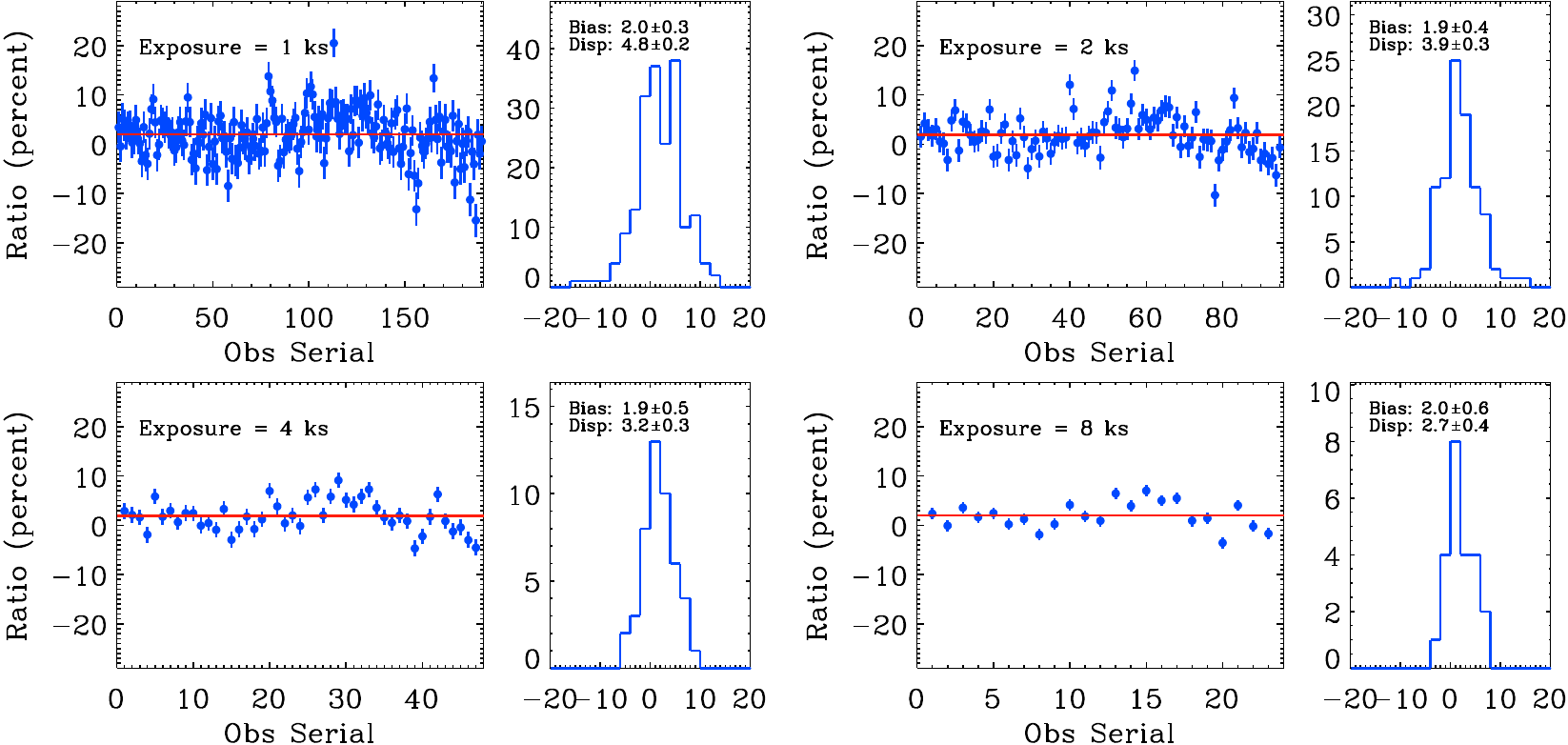}}
\caption{Distributions of the residuals in the background spectra estimation in the 200th channel with exposures 1~ks, 2~ks, 4~ks, and 8~ks.
In each panel, the broadening of the histogram is the dispersion of the residuals that caused 
by both the statistical error of the test data and the systematic error of the background model.}
\label{Fig:err_sys_200th}
\end{figure*}

\begin{figure*}
\center{
\includegraphics[angle=0,scale=1.00]{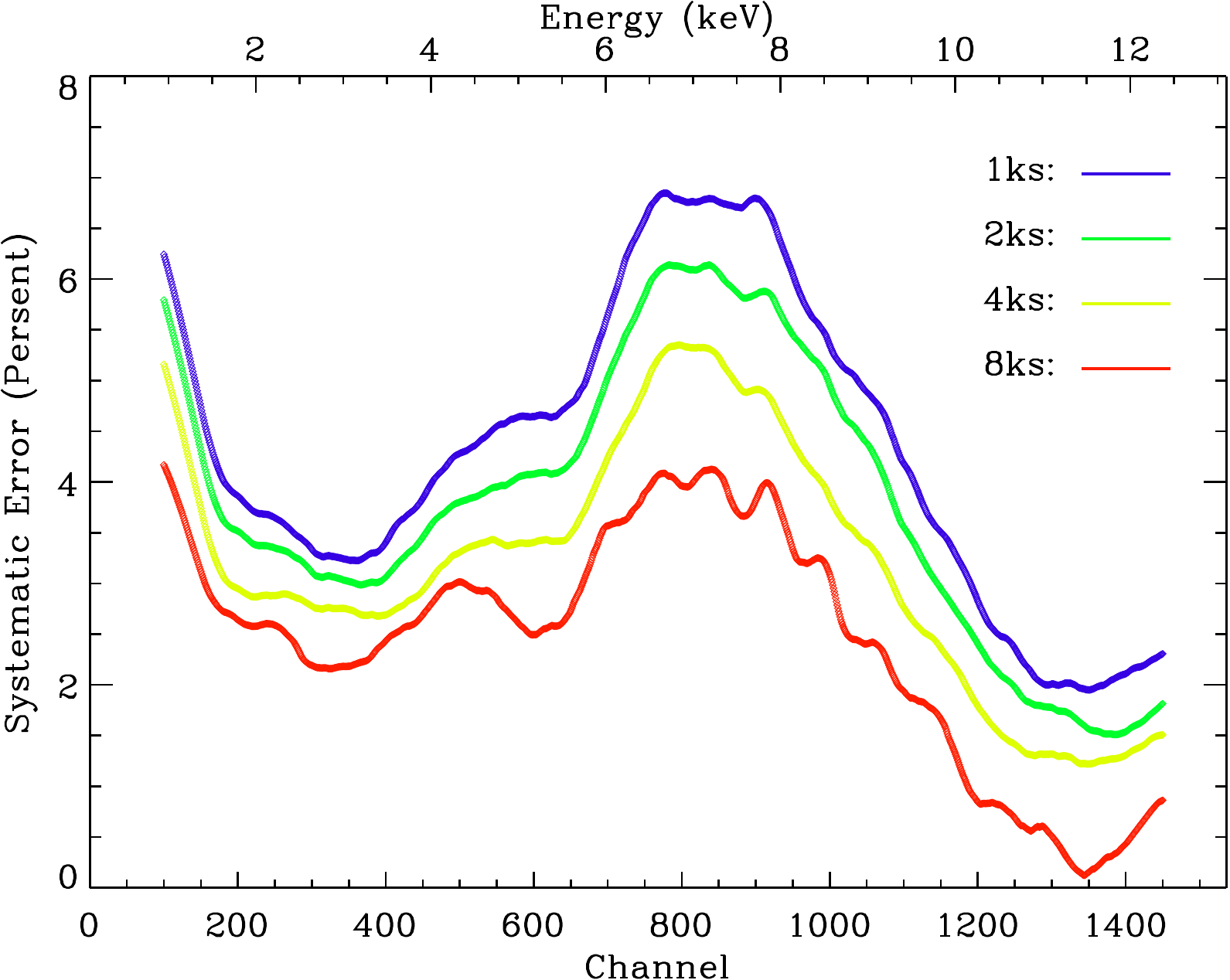}}
\caption{Systematic errors of the background spectra estimations in $1-12.4$~keV with the exposures 1~ks, 2~ks 4~ks, and 8~ks.}
\label{Fig:err_sys_spec_channel}
\end{figure*}

The same processes are also performed for $T_{\rm exp} = 2,4$ and 8~ks to obtain the systematic errors specific to these exposures. 
Figure~\ref{Fig:err_sys_200th} shows the distributions of the residuals resulted in the 200th channel of the background modeling with exposures 1 ks, 2 ks, 4 ks, and 8 ks, respectively.
In each panel, the broadening of the histogram is the dispersion of the residuals caused by both 
the statistical error of the test data and the systematic error of the background model.
As shown in Figure~\ref{Fig:err_sys_spec_channel} and Table~\ref{tab:systematic_error_spec}, 
the systematic error is 3.7\% \& 2.0\% for $T_{\rm exp} = 1$ \& 8~ks in $1-10$~keV, respectively. Because the systematic error 
is partly propagated from the statistical errors of the blind FOV detectors, which is reduced as the exposure increases, 
there is an anti-correlation between the systematic error and the exposure.

\subsection{Test for the Background Light Curve}
\begin{figure*}
\center{
\includegraphics[angle=0,scale=1.00]{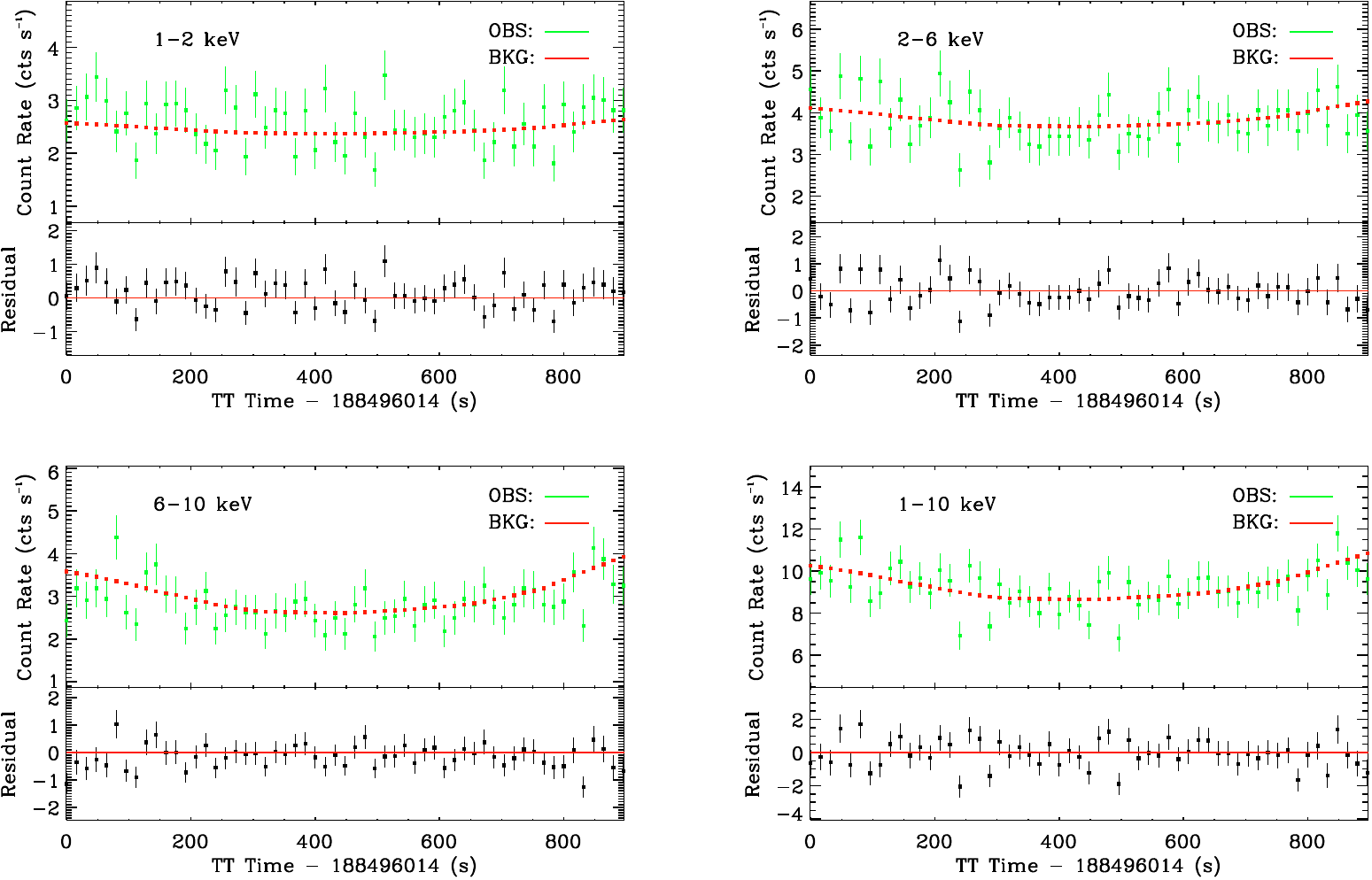}}
\caption{Test of the background light curve estimation in $1-2$~keV,  $2-6$~keV, $6-10$~keV and $1-10$~keV with $T_{\rm bin}=16~{\rm s}$. 
For each panel, the observed (green) and estimated (red) background light curves are shown in the top, and the residual is shown in the bottom.}
\label{Fig:bkg_lc_test}
\end{figure*}

With the process described in Section 4.2, the background light curve can be obtained (Figure~\ref{Fig:bkg_lc_test}).
All the blank sky observations are processed to obtain the light curves with four energy bands and eight time bins (Table~\ref{tab:systematic_error_lc}). 
In the following, the background light curve in $1-10$~keV with $T_{\rm bin} = 1~{\rm s}$ is used as an example to shown the 
detailed processes of the model test.

(1) Merge all the blank sky observations into one observation with the energy band $1-10$~keV and $T_{\rm bin} = 1~{\rm s}$ 
and obtain the observational light curve ($L_{\rm obs}$).

(2) Obtain the particle background light curve in $1-10$~keV by the particle background model.

(3) Combine the particle background light curve and the diffuse X-ray background $1-10$~keV in high Galactic latitude to 
obtain the total background light curve ($L_{\rm bkg}$).

(4) Compare $L_{\rm obs}$ and $L_{\rm bkg}$ to obtain the residuals $R_{\rm L} = {L_{\rm obs} - L_{\rm bkg}}$.

(5) Calculate systematic deviation and uncertainty with the method described in Section 5.2.

\begin{figure*}
\center{
\includegraphics[angle=0,scale=1.00]{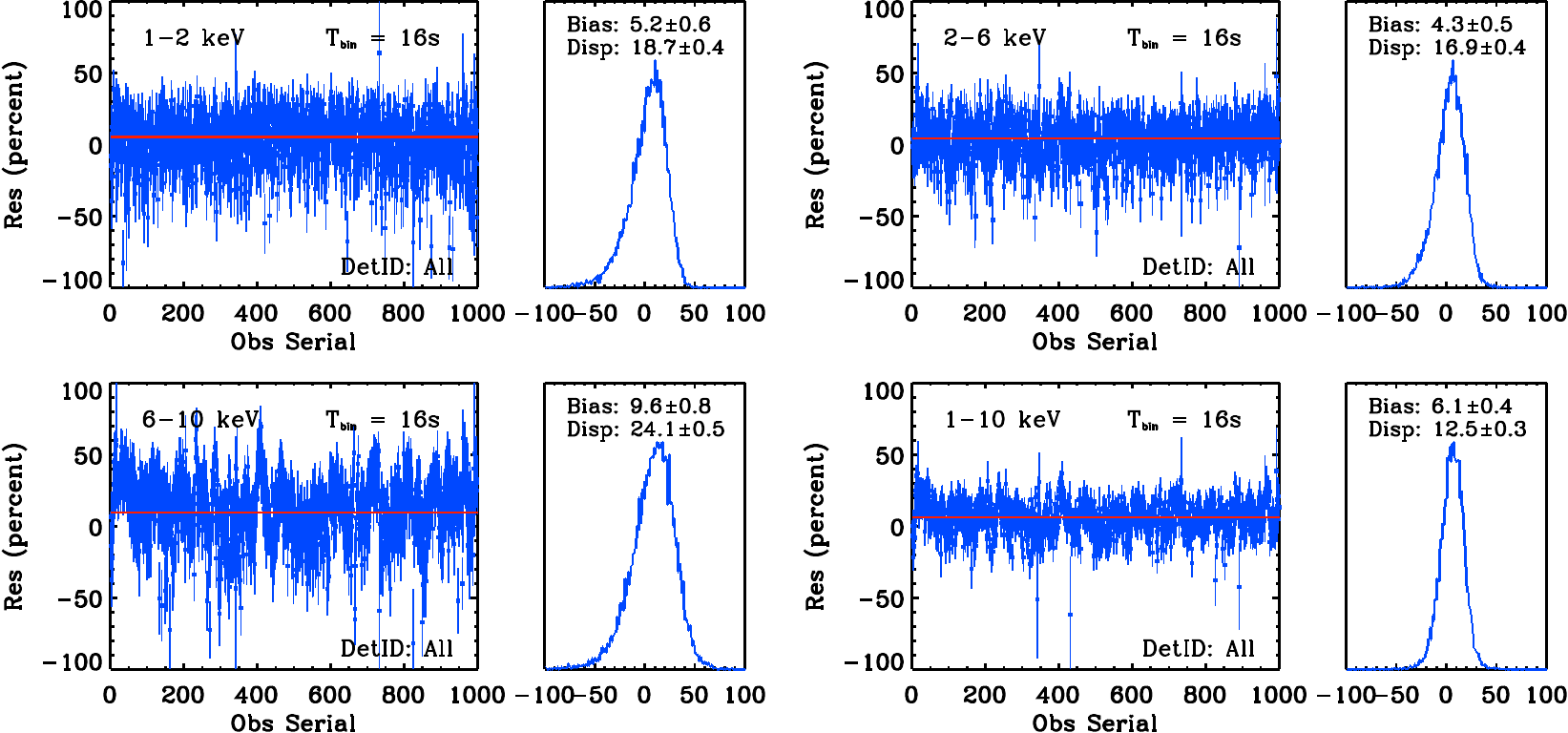}}
\caption{Distributions of the residuals in the background estimations of the light curves with $T_{\rm bin}=16$~s for four energy bands.
In each panel, the broadening of the histogram is the dispersion of the residuals that caused 
by both the statistical error of the test data and the systematic error of the background model.}
\label{Fig:err_sys_lc_16s}
\end{figure*}

The above steps are also performed with the other parameters shown in Table~\ref{tab:systematic_error_lc}. 
The distributions of the residuals with $T_{\rm bin} = 16~{\rm s}$ in four energy bands are shown in Figure~\ref{Fig:err_sys_lc_16s}, 
and the detailed results of the systematic error analyses are also shown in Table~\ref{tab:systematic_error_lc}.

\section{SUMMARY}
More than 150 blank sky observations are performed to study the background of \emph{Insight-HXMT}/LE.
We developed a new method to optimize the GTI selection, with which the abnormal flares can be removed 
in the preliminary data reduction for every blank sky observation. Both the in-orbit observation and 
the on-ground simulation indicate that the background of the small FOV detector of \emph{Insight-HXMT}/LE 
is mainly composed of two parts; one is the particle background that is dominant in high-energy band 
($>7$~keV), and another is the diffuse X-ray background that is dominant in low-energy band ($<7$~keV). 
With the blank sky observation, we find two important characteristics of the particle background, i.e., 
the spectral shape are consistent in every geographical location and the intensity can be indicated by 
the contemporary flux of the blind FOV detector in $10-12.5$~keV. Based on these two characteristics, 
we developed the method to estimate the particle background of \emph{Insight-HXMT}/LE. In addition, the 
diffuse X-ray background caused by the CXB is also obtained with the blank sky observation in the high 
Galactic latitude region. We investigated the exposure dependence of the systematic error induced from
the current background modeling. For $T_{\rm exp} > 1$~ks, the systematic error of the estimated background 
spectra is $\sim3\%$ in $1-10$~keV on average, which are partly attributed to the systematic error
propagated from the statistical error of the blind detector during the background estimation. We also
make the analysis to the light curve with different time bins. For the background light curves in $1-10$~keV, 
the systematic errors are in general $<8\%$. It is worth noting that the diffuse X-ray background in the 
Galactic center can be 2 times higher than that in the high Galactic region. Therefore, for the pointing 
observation of the Galactic center, the diffuse X-ray background is obtained from the sky map of the LE 
diffuse X-ray background, which will be described elsewhere in a forthcoming paper and thus not reported here.

\acknowledgments
This work made use of the data from the \emph{Insight-HXMT} mission, a project funded by China National Space Administration (CNSA) 
and the Chinese Academy of Sciences (CAS). The authors thank supports from the National Program on Key Research and Development Project 
(Grant No. 2016YFA0400802 and 2016YFA0400801) and the National Natural Science Foundation of China under Grants No. U1838202 and U1838201.

\renewcommand{\arraystretch}{1.2}
\begin{deluxetable}{ccccc}
  \tablewidth{0pt}
  \tablecaption{Systematic errors of the background spectra ($1-10$~keV) with different exposures}
  \tablehead{  $T_{\rm exp}$  &  1~ks  &  2~ks  &  4~ks  &  8~ks  }
  \startdata
      $\sigma_{\rm sys,~1-7~keV}$   &  4.2\%  &  3.8\%  &  3.3\%  &  2.8\%  \\
      $\sigma_{\rm sys,~1-10~keV}$  &  3.7\%  &  3.2\%  &  2.7\%  &  2.0\%
	  \enddata
	  \label{tab:systematic_error_spec}
\end{deluxetable}

\renewcommand{\arraystretch}{1.2}
\begin{deluxetable}{ccccccccc}
  \tablewidth{0pt}
  \tablecaption{Systematic errors of the background light curves with different time bins}
  \tablehead{  Energy Band  &  16~s  &  32~s  &  64~s  &  128~s  }
  \startdata
	  $1-2$ keV   &  6.2\%  &  5.8\%  &  5.9\%  &  5.9\% \\ 
	  $2-6$ keV   &  6.0\%  &  6.2\%  &  6.2\%  &  6.2\% \\ 
	  $6-10$ keV  & 12.7\%  & 13.1\%  & 13.3\%  & 13.5\% \\ 
	  $1-10$ keV  &  7.6\%  &  7.6\%  &  7.7\%  &  7.8\% \\ 
	  \enddata
	  \label{tab:systematic_error_lc}
\end{deluxetable}


%

%
%
%
%

\end{document}